\documentstyle[epsfig,aps]{revtex}
\begin{document}
\draft
\title{Slow light propagation in trapped atomic quantum gases}
\author{\"Ozg\"ur E. M\"ustecapl{\i}o\u{g}lu and L. You}
\address{
School of Physics, Georgia Institute of Technology, Atlanta GA
30309-0432, USA}
\date{\today}
\maketitle
\begin{abstract}
We study semi-classical slow light propagation in trapped two level atomic
quantum gases. The temperature dependent behaviors of both group velocity
and transmissions are compared for low temperature Bose, Fermi, and Boltzman
gases within the local density approximation for their spatial density profile.
\end{abstract}
\narrowtext
Recently, a dramatic demonstration of slow light propagation,
down to $17$ (m/s) was reported \cite{hau,inouye}. Achievement of
such an extremely low speed owes mostly to the widely discussed
phenomenon of electromagnetically induced transparency (EIT),
which makes propagation of light in an otherwise opaque medium
possible \cite{harris}. In related theoretical and
experimental studies, three level atomic vapours are the
typical medium. Although Bose-Condensation is not crucial
to EIT or slow light propagation \cite{hau,inouye}, the long
coherence time of such a degenerate quantum medium does
prove to be advantageous. Similar observations have since been
carried out in optically dense hot rubidium
vapours [$\sim 90$ (m/s)] \cite{kash} and in a
optically thick Pr:YSO crystal [$\sim 45$ (m/s)] \cite{turukhin}.
Possibilities of vanishing or even negative group velocities, in
a coherently driven Doppler broadened atomic vapours have also
been discussed\cite{kocharovskaya}.

Ultra-slow light propagation offers many potential applications
since macroscopically the medium can be viewed as possessing a
high index of refraction, albeit within a narrow spectrum.
The accompanied super-high nonlinear coupling between weak fields
in a long coherence medium such as a Bose-Einstein condensate (BEC)
opens up novel regimes of quantum nonlinear optics \cite{harris-hau}.
Application to quantum networks and quantum information
processing, including quantum entanglement of slow
photons\cite{lukin-imamoglu}, non-classical (e.g. squeezed) and
entangled atomic ensembles\cite{lukin-yelin}, quantum
memories\cite{fleischhauer-lukin}, have been proposed.
Their implications to quantum non-demolishing measurements and high
precision spectroscopy using squeezed light have been suggested
through enhanced acousto-optical effects\cite{matsko} and
narrow-band sources for non-classical radiation \cite{fleischhauer-lukin-matsko}.


The aim of this paper is two fold: (1) Inspired by the recent
theoretical modeling of slow light propagation \cite{hau} in BEC
by Morigi and Agarwal \cite{morigi}, we ask the question of
comparative differences related to spatial density profiles of
different quantum gases; (2) We explore a simpler model composed
of two level atoms \cite{you}. In our formulation, we consider the
semi-classical propagation of a laser pulse through an ultra-cold
quantum gas of two level atoms described by a density profile
$\rho(\vec{r},t)$. At very cold temperatures atoms are highly
delocalized and continuum treatment of the gas as a medium is
desirable. The polarization density operator is thus
$\vec{P}(\vec{r},t)=\vec{d}_{ge}\psi_g^{\dag}(\vec{r},t)
\psi_e(\vec{r},t)+{\rm h.c.}$ \cite{krutitsky_1} with
$\vec{d}_{ge}$ the electronic dipole transition moment between
ground state $|g\rangle$ and excited state $|e\rangle$.
$\psi_{\eta=g,e}(\vec{r},t)$ is the second quantized field
operator for atoms in state $|g\rangle$ and $|e\rangle$
respectively. An incident laser pulse polarizes the medium and
the induced dipoles of atoms subsequently emit coherent radiation.
For most of aforementioned applications, the atomic density
profile remains an essential ingredient. In addition, careful
determination of the optical response inside a dense medium
requires the inclusion of local field corrections. In case of two
level atoms as studied here, it is necessary to also have a
significant transmission through the medium. At any point within
the medium, the electric field is given by the superposition of
the incident field and the radiated field, the latter excludes
the self-field radiated by the dipole at this same point.
Adopting the standard approach, we write the total local field
acting on the atoms
 $\vec{E}_L=\vec{E}_P+\vec{E}_D+\vec{E}_V$,
as a sum of the incident field ($\vec{E}_P$),
the field generated by other atomic dipoles
($\vec{E}_D$) and the field due to vacuum fluctuations
($\vec{E}_V$). The dipole field depends on the type of medium
as well as its density profile.

Following standard procedures \cite{allen-eberly} we eliminate
fast radiation field dynamics. It is then possible to recognize
that $\vec{E}_D$ is given by $[\vec{P}_t]$, the retarded
transverse polarization field \cite{krutitsky_1,jackson}. We
obtain
\begin{eqnarray}\label{dipole-field}
  \vec{E}_D(\vec{r},t) =
  \vec{\nabla}\times\vec{\nabla}\times\int_{V/\{\vec{r}\}}
  d\vec{r}^{\,\prime}\frac{\vec{P}(\vec{r}^{\,\prime},
  t-|\vec{r}-\vec{r}^{\,\prime}|/c)}{|\vec{r}-\vec{r}^{\,\prime}|},
  \nonumber
\end{eqnarray}
where $V/\{\vec{r}\}$ indicates that integral over the interaction
region excludes a small region around $\vec{r}$.
Proper evaluation of the integral in the vicinity of $\vec{r}$
yields the Lorentz-Lorenz
correction (local field correction)\cite{jackson},
\begin{eqnarray}\label{lorentz-lorenz}
  \vec{E}_{L}=\vec{E}+\frac{4\pi}{3}\vec{P}.
\end{eqnarray}
The field $\vec{E}$, including contributions of all dipoles in the
interaction volume $V$, is the macroscopic field governed by
the Maxwell equations. It is important to point out that the
Lorentz-Lorenz relation Eq. (\ref{lorentz-lorenz}) holds true
for both linear and nonlinear media \cite{bowden}.
To maintain sufficient transmission, we assume the
incident field (of central frequency $\omega$)
to be far-off resonant (from atomic resonance $\omega_0$)
with a large detuning $\Delta=\omega-\omega_0$.
When the incident pulse is weak,
the linear response theory probes the unperturbed density profile
$\rho(\vec{r},t)\approx
\rho(\vec{r})=\psi_g^\dag(\vec{r},0)\psi_g(\vec{r},0)$.
For stronger higher input fields, the density of the gas needs to
be determined self-consistently \cite{wallis}.
The linear optical response function
(the electric susceptibility, $\chi$)
follows the Clausius-Mossotti relation
\begin{eqnarray}\label{clausius-mossotti}
 \chi(\vec{r})
 =\frac{\alpha\rho(\vec{r})}{1-\frac{4\pi}{3}\alpha\rho(\vec{r})
 +i\frac{\gamma}{2\Delta}},
\end{eqnarray}
involving Lorentz-Lorenz shift as a first correction
\cite{krutitsky_1,wallis,morice,ruostekoski_1,ruostekoski_2,fleischhauer-yelin,krutitsky_2}.
$\alpha=|\vec d_{ge}|^2/\hbar\Delta$ is the approximate
off-resonant atomic polarizability and $\gamma$ is the excited
state spontaneous emission rate. Higher order quantum corrections
to this formula arise from multiple scattering of photons within
neighboring atomic pairs as well as from many-body correlations.
These effects are neglected in this study for the coherent
propagation effect. Recent studies of these corrections found the
weak field optical response Eq. (\ref{clausius-mossotti}) works
well in available experimental density regimes for both Bose
\cite{krutitsky_1,wallis,krutitsky_2} and Fermi gases
\cite{krutitsky_2,ruostekoski_3}. The main physical reason
allowing for such a simplification is the experimental effort
in assuring a low optical density $\chi(\vec{r})=\alpha\rho(\vec{r})$
to provide reliable probes of the system.
Local field correction is significant
when the number of atoms in a characteristic volume
\begin{eqnarray}
V_{\alpha}=\frac{4\pi}{3}\alpha=4\pi^2\frac{\gamma}{\Delta k_L^3},
\end{eqnarray}
is comparable to unity. Here, $k_L$ is the wave number of the
near-resonant incident field with detuning a $\Delta$
($\approx 10\gamma$ in this study)
from the transition wavelength $\lambda=589$ (nm). This gives
$V_{\alpha}\approx 4\times 10^{-15}\,{\rm cm}^{3}$.
Hence, only for densities around
$10^{14}\,{\rm cm}^{-3}$ or higher, will local field start to
cause qualitative/quantitative modifications to the optical
response of an atomic cloud. For a non-interacting
condensate of $10^6$ atoms, it is estimated \cite{morigi}
using trap parameters of Ref. \cite{hau} that an equivalent
homogeneous density is $\rho\approx 8\times 10^{15}\,{\rm cm}^{-3}$.
The presence of a repulsive interaction between atoms
lowers that value to $\rho\approx
3\times10^{13}\,{\rm cm}^{-3}$ (using Thomas-Fermi approximation).
Therefore, it is more serious to
neglect local field effects in a BEC model of
non-interacting atoms than our semi-classical treatment for
an interacting condensate. Although the above estimates
could justify the neglect of Lorentz-Lorenz shift,
we decided to keep it for a couple of reasons.
First, these estimates rely on a homogeneous density profile.
Yet, the density of an actual condensate can be much higher
near the trap center. For an inhomogeneous condensate
the group velocity is determined by a spatial averaged optical
response, it is therefore necessary to check consistently any
deviations from the column density approach used for a homogeneous
density distribution.
Second, considerable arguments (see discussions in next paragraph)
point to the fact that local field effects, many body correlations, and multiple
scattering effects are of the same order. Keeping Lorentz-Lorenz shift
thus allows for a consistent check against approximations in neglecting
the many body (e.g. second order density-density) correlations as
well as multiple scattering.

It is a challenging task to include quantum many-body correlations
and multiple scattering of the medium into an effective optical
response (effective refractive index). In an earlier attempt, a
perturbative density expansion, similar to the Virial expansion,
was employed in Ref. \cite{morice} to calculate the first two
corrections. The effect was found to be negligible for atomic
densities $\ll k_L^3$. Later, it was shown that such an
approximation is equivalent to truncating an infinite hierarchy of
equations for correlation functions
\cite{ruostekoski_1,ruostekoski_2,ruostekoski_3}. Furthermore, it
was shown under the truncation approximation that correlation
correction terms are of the same order as the local field
corrections. A notable exception is the low density Fermi gas
in a Cooper paired BCS state \cite{ruostekoski_4}, a case not
considered here. The neglect of quantum correlations even in
the ideal fermi gas case as studied here may affect some
of our results. Other recent studies include self-consistent
analysis in \cite{krutitsky_1,wallis,krutitsky_2} and a Dyson
equation formulation in \cite{fleischhauer-yelin}.

The real part of the susceptibility ($\chi^{\,\prime}$) is related
to the index of refraction from which we can calculate the group
velocity according to \cite{kash}
\begin{eqnarray}\label{v-group-defn}
  v_g(\omega,\vec{r}) \approx {c\over
  1+2\pi\chi^{\,\prime}+2\pi\omega
  \frac{\partial\chi^{\,\prime}}{\partial\omega}},
\end{eqnarray}
where $c$ is the speed of light in vacuum, and the probe pulse
central wave number is $k=\omega/c$.
In obtaining the above equation, we have neglected a small
contribution from $\partial\chi^{\,\prime}/\partial k$,
and assumed the
susceptibility changes slowly over an optical wavelength $2\pi/k$.
The result is,
\begin{eqnarray}\label{v-group}
  v_g = c\left(1+\frac{2\pi\omega_0\alpha\rho}
  {\Delta(1-4\pi\alpha\rho/3)^2}\right)^{-1}.
\end{eqnarray}

In the reported measurements of slow group velocity
\cite{hau,inouye}, two separate parameters are needed:
1) the effective propagation length of the medium;
2) the delay time of the probe pulse.
The ratio of these two quantities defines the effective
group velocity \cite{hau}. We adopted the definitions
as used by Morigi and Agwaral \cite{morigi} for a
three level Bose gas, and appropriately generalized
to the case of an interacting Bose gas. We then compare
with corresponding results for an (ideal) Fermi and
Boltzman gas.

The ground state atomic density profile for a Fermi gas is
obtained following the semi-classical approximation given by
\begin{eqnarray}\label{fdd}
  \rho_F(\vec{r})=\int
  \frac{d\vec{p}}{h^3}\frac{1}{e^{\beta[H(\vec{p},\vec{r})-\mu]}+1},
\end{eqnarray}
where $\beta=1/(k_BT)$. The Hamiltonian $H(\vec{p},\vec{r})$
describes ground state atomic motion inside an external trapping
potential $V(\vec{r})=M\omega_r^2(r^2+\epsilon^2 z^2)/2$.
$\omega_r$ is the radial trap frequency, $\epsilon$ is the aspect
ratio, and $M$ is the atomic mass. The probe light is assumed to
propagate along the (long) $z-$axis of the atomic cloud. The
chemical potential $\mu$ is determined by the normalization
$N=\int d\vec{r}\rho_F(\vec{r})$ with $N$ the total number of
atoms. In the low temperature limit, $\mu$ can be described
within the Sommerfeld expansion, while the high temperature
behavior is found by direct integration similar to a Boltzmann gas
\cite{butts},
\begin{eqnarray}\label{mu-fermi}
  \mu = \left\{\begin{array}{ll}
             -\ln{[6(T/T_F)^3]}/\beta, & T>0.55T_F, \\
             E_F[1-\pi^2(T/T_F)^2/3], & T\leq 0.55T_F,
             \end{array}\right.
\end{eqnarray}
where $T_F$ and $E_F$ are the Fermi temperature
and Fermi energy.

For an interacting Bose gas, the ground state
density profile is computed using the analytic fitting
function as developed in \cite{naraschewski}. Below the
condensation temperature $T_C$, the condensate
component is described by the Thomas-Fermi approximation
\cite{naraschewski,bagnato},
\begin{eqnarray}\label{bec}
  \rho_B(\vec{r}) =
  \frac{\mu-V}{U}
  \theta(\mu-V)\theta(T_c-T)
  +\frac{g_{3/2}({\sf z} e^{-\beta V})}
  {\Lambda_T^3},
\end{eqnarray}
with $T_c$ is the critical temperature.
$U=4\pi\hbar^2 a_{\rm sc}/M$ with $a_{\rm sc}$
the atomic scattering length. $\theta(.)$ is the Heaviside step
function, $g_{n}(x)=\sum_j x^j/j^{n}$, and $\Lambda_T$ is
the thermal de Br\"oglie wavelength. The chemical potential at
high temperatures is determined by solving
${\rm Li}_3({\sf z})=(T/T_c)^{-3}\zeta(3)$ in terms of the
fugacity ${\sf z}=e^{\beta\mu}$. ${\rm Li}_3(.)$ and
$\zeta(3)$ are the third order polylogarithmic and
Riemann-Zeta functions respectively.
At low temperatures, it is found that \cite{naraschewski},
\begin{eqnarray}\label{mu-bec}
  \mu=\mu_{TF} \left( \frac{N_0}{N} \right)^{2/5},
\end{eqnarray}
with $\mu_{TF}$ the Thomas-Fermi approximation chemical potential.
The condensate fraction is then given by
\begin{eqnarray}\label{cond-frac}
 \frac{N_0}{N} =
 1-\left(\frac{T}{T_c}\right)^3-\eta\frac{\zeta(2)}{\zeta(3)}\left(\frac{T}{T_c}
 \right)^2\left[1-\left(\frac{T}{T_c}\right)^3\right]^{2/5},
\end{eqnarray}
with a scaling parameter $\eta$ defined as
\begin{eqnarray}\label{scaling}
  \eta =
  \frac{\mu_{TF}}{k_BT_c}=\frac{1}{2}\zeta(3)^{1/3}\left(15N^{1/6}\frac{a_{\rm sc}}
  {a_{ho}}\right)^{2/5}.
\end{eqnarray}
$a_{\rm ho}=\sqrt{\hbar/(M\epsilon^{1/3}\omega_r)}$ is the average
harmonic oscillator length scale \cite{giorgini}.

Finally, for a classical gas we use
\begin{eqnarray}\label{boltzmann}
  \rho_C(\vec{r}) = \frac{1}{\Lambda_T^3}e^{-\beta(V-\mu)},
\end{eqnarray}
with the chemical potential $\mu=-\ln{[6(T/T_F)^3]}/\beta$.

\begin{figure}[t]
\centerline{\epsfig{file=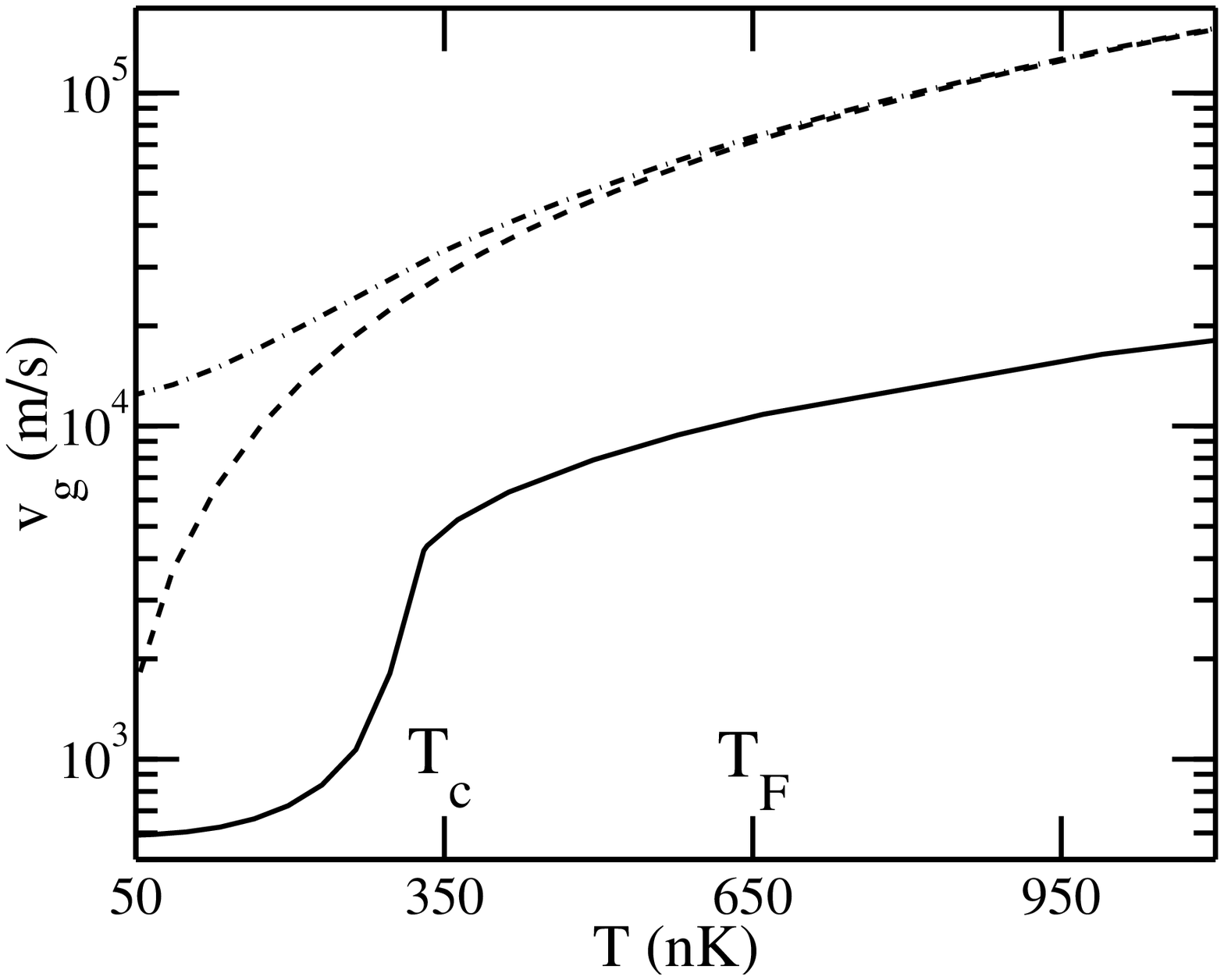,width=7.cm}\\[12pt]}
\caption{Temperature dependent group velocity in Fermi
(dash-dotted line), Bose (solid line), and Boltzman
(dashed line) for two level trapped quantum gases.
We used Na as an example with the following
parameters:
$N=3.8\times10^6$, $R=7.5$ ($\mu$m),
$\omega_r=(2\pi)\, 69$ (Hz),
$\epsilon=1/3$, $\gamma=(2\pi)\, 10.03$ (Hz),
$a_{\rm sc}=2.75$ (nm),
$\omega_0=(2\pi)\, 5.1\times10^{14}$ (Hz)
as in Ref. [1]. For these parameters $R_B=17.76$ ($\mu$m),
$R_F=50.04$ ($\mu$m), $a_r=2.52$ ($\mu$m),
and $a_{\rm ho}=3.03$ ($\mu$m).
We chose the detuning $\Delta=10\gamma$
to assure sufficiently transmission.}
\label{fig1}
\end{figure}

From each of the above three distributions for $\rho(\vec r)$,
we first determine the effective length of the medium
using
\begin{eqnarray}\label{size}
  L(T) = \left[\int_V d\vec{r} z^2 \rho(\vec{r})\right]^{1/2}.
\end{eqnarray}
In a typical experiment, a pinhole (of radius $R$ less than the
Thomas Fermi radius) is often introduced just after the atomic
cloud to restrict delay time detection to only light passing
through the most dense central axial column.
This can be included by the averaged delay time
\begin{equation}\label{group-delay}
t_d(T) = \frac{1}{\pi R^2}\int_0^R 2\pi dr
\int_{-L}^{L}dz\frac{1}{v_g(\vec{r})}-\frac{2L}{c}.
\end{equation}
For small pinhole size and long optical path length $L$, we can
ignore the spatial dependence of $L$ due to shape of the cloud
\cite{morigi}. The group speed is then calculated according to
$v_g = L(T)/t_d(T)$. Typical reports of our calculation are
reported in Fig. \ref{fig1}. Below the BEC transition temperature
we see that the group velocity in a Bose gas drops sharply.
Deviations of Fermi gas from the classical behavior, on the other
hand, appear below half of $T_F$. For the parameters used
$N=3.8\times 10^6$ \cite{hau} and other parameters as given in
Fig. \ref{fig1} caption, $T_F$ \cite{Jin} is approximately twice
the $T_c$. In such degenerate low temperature regimes, different
quantum statistics of the gases result in qualitatively and
quantitatively different behaviors of their slowed group
velocities. For this set of experimental parameters, we found
local field correction causes only $\sim3\%$ reduction of group
velocity, invisible in the figure.

\begin{figure}[t]
\centerline{\epsfig{file=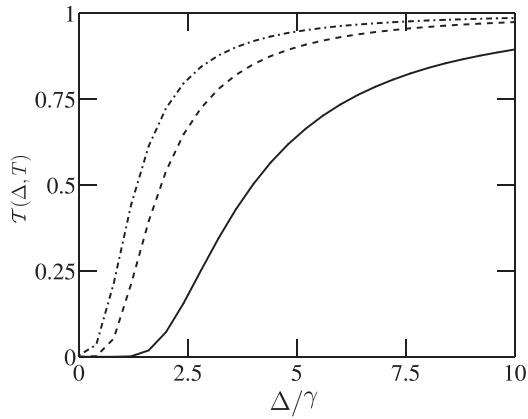,width=7.cm}\\[12pt]}
\caption{The same as in Fig. 1, except for the
detuning dependence of the transmission
coefficient. All three quantum gases are compared for
the same temperature $T=T_c/2 \approx T_F/4$.} \label{fig2}
\end{figure}

The absorption caused by
the imaginary part of the susceptibility
($\chi^{\,\prime\prime}$) is best described by the
the transmission coefficient
\begin{eqnarray}\label{alpha}
  \alpha_{\cal T} = -2\frac{\omega_0}{c}\frac{1}{\pi R^2}
  \int_0^R2\pi r dr \int_{-L(T)/2}^{L(T)/2} dz
  \chi^{\,\prime\prime},
\end{eqnarray}
which in turn gives the transmission ${\cal
T}(\Delta,T)=e^{\alpha_{\cal T}}$. A good estimate is provided by
${\cal T}=e^{-2\omega_0\chi_{m}^{\,\prime\prime}L(T)}$, with
$\chi_{m}^{\,\prime\prime}$ the imaginary part at the peak
density (at $\vec r=0$). Our numerical results are presented in
Fig. \ref{fig2} for $T\sim T_c/2\approx T_F/4$, in the quantum
degenerate regime. We notice the transmission for a Bose gas is
always lowest, while for the corresponding Fermi is always the
highest. This can be simply explained in terms if their peak
densities. For near-resonant pulse propagation with
$|\Delta|<3\gamma$ Bose gas becomes essentially opaque, while
significant transmission through a Fermi gas is still possible.
This high transmission feature can be considered as an advantage
of a Fermi gas, over the Bose gas regarding applications in quantum
memories, and optical data storage.

It is interesting to discuss briefly results for the zero
temperature limit, in which analytical solutions exist
for all density distribution functions, given respectively by
\begin{eqnarray}\label{zero-density}
  \rho_F(T=0) &=&
  \frac{8N\epsilon}{\pi^2R_F^6}(R_F^2-r^2-\epsilon^2z^2)^{3/2},\nonumber\\
  \rho_B(T=0) &=& \frac{15N\epsilon}{8\pi
  R_B^5}(R_B^2-r^2-\epsilon^2z^2),\nonumber
\end{eqnarray}
where
$R_B=(15N\,\epsilon\, a_{\rm sc}/a_{\rm ho})^{1/5}a_r$
and
$R_F=(48N\epsilon)^{1/6}a_r$
are the Thomas-Fermi radius for bosonic and fermionic clouds
and $a_r=\sqrt{\hbar/(M\omega_r)}$.

After some tedious calculations, we obtain
\begin{eqnarray}
v_g =
   \frac{4\omega_0^2\Delta^2}{3\sqrt{7}N\epsilon c^2\gamma}
   R^2R_B\left(1-\left[1-\left(\frac{R}{R_B}\right)^2\right]^{5/2}\right)^{-1}\nonumber
\end{eqnarray}
for bose gas and
\begin{eqnarray}
   v_g= \frac{\sqrt{2}\omega_0^2\Delta^2}{9N\epsilon
   c^2\gamma}R_F^3\left[1-\left(\frac{R}{R_F}\right)^2+\frac{1}{3}
   \left(\frac{R}{R_F}\right)^4\right]^{-1} \nonumber
\end{eqnarray}
for fermi gas respectively at $T=0$..

Valuable insight can be gained by examining the external parameter
dependence of the above two formulae. Most notably $v_g \sim N^{-2/5}$ for a
Bose gas while $v_g \sim N^{-1/2}$ for a Fermi gas. Larger
delay times are obtained for $R \ll R_F, R_B$ than for $R\sim
R_F, R_B$, and depends on the statistical nature of the atomic medium.
In a Fermi gas, $v_g$ is three times larger for $R\sim R_F$ than for $R\ll
R_F$. While in the case of Bose gas, $v_g$ is $2.5$ times larger
for $R\sim R_B$ than for $R\ll R_B$. This dependence may be
used to calibrate and determine experimental observables such as
the cloud size, temperature, and the Thomas-Fermi radius.

In conclusion, we have examined the temperature dependent
behavior of slow light transmission through an ultra-cold
quantum gas of two level atoms. We have included local field
corrections, and compared the results
for atoms with different quantum statistics.
Analytical results are obtained below the
condensation and Fermi temperatures.
It is interesting to point out that as displayed in Fig.
\ref{fig1}, the logarithmic of group velocity for a Fermi gas
has a simple temperature dependence; it is almost
linear. This feature may be used for precise temperature
measurements in regimes below the Fermi temperature
when precise calorimetry is difficult to achieve.
The absolute scale can be obtained by a calibration
against other independent measurements at higher temperatures.
Furthermore, as emphasized in the theoretical studies
in Ref. \cite{morigi}, a sharp discontinuity shows up
for a Bose gas exactly at the condensation phase
transition point. This is clearly displayed in Fig. \ref{fig1}.
For a three-level system, theoretical results for $T<T_C$
are less reliable because of complications from both
strong mean-field mutual couplings and multiple
scattering effects. Consequently, agreement with experiment
in this regime is poor \cite{morigi}. In the two-level case
studied here, our theory includes the self-mean-field
interaction, and one may expect its validity even
in this ultra-low temperature regime as long as the
incident probe field is weak. Unfortunately, there
is no current experimental data to compare with. Should this
prove to be true, this dramatic feature at the transition
temperature can be used for accurate determination of $T_C$.
Finally we note that although these results are
similar to those in three level EIT systems.
The two level model studied here may present some
practical experimental advantages.

We acknowledge insightful discussions with Dr. P. Zhou.
This work is supported by the NSF grant No. PHY-9722410..


\end{document}